\documentclass[%
aps,
pre,
 amsmath,amssymb,
 superscriptaddress,
preprint,%
longbibliography
]{revtex4-1}

\usepackage{graphicx}
\usepackage{dcolumn}
\usepackage{bm}
\usepackage{ulem}
\usepackage[utf8]{inputenc}
\usepackage[T1]{fontenc}
\usepackage{mathptmx}
\usepackage{textcomp} 
\usepackage{nicefrac}
\usepackage{xcolor}
\usepackage{gensymb}
\usepackage{multibib}

\usepackage{lineno}

\DeclareMathAlphabet{\pazocal}{OMS}{zplm}{m}{n}

\begin{document}


\title[Ultrafast dynamics of bright and dark excitons\\ in monolayer WSe$_2$ and heterobilayer WSe$_2$/MoS$_2$]{Ultrafast dynamics of bright and dark excitons\\ in monolayer WSe$_2$ and heterobilayer WSe$_2$/MoS$_2$}






\author{Jan Philipp Bange} %
\address{I. Physikalisches Institut, Georg-August-Universit\"at G\"ottingen, Friedrich-Hund-Platz 1, 37077 G\"ottingen, Germany}

\author{Paul Werner} %
\address{I. Physikalisches Institut, Georg-August-Universit\"at G\"ottingen, Friedrich-Hund-Platz 1, 37077 G\"ottingen, Germany}

\author{David Schmitt} %
\address{I. Physikalisches Institut, Georg-August-Universit\"at G\"ottingen, Friedrich-Hund-Platz 1, 37077 G\"ottingen, Germany}

\author{Wiebke Bennecke} %
\address{I. Physikalisches Institut, Georg-August-Universit\"at G\"ottingen, Friedrich-Hund-Platz 1, 37077 G\"ottingen, Germany}

\author{Giuseppe Meneghini} 
\address{Fachbereich Physik, Philipps-Universit{\"a}t, 35032 Marburg, Germany}

\author{AbdulAziz AlMutairi} 
\address{Department of Engineering, University of Cambridge, Cambridge CB3 0FA, U.K.}

\author{Marco Merboldt} %
\address{I. Physikalisches Institut, Georg-August-Universit\"at G\"ottingen, Friedrich-Hund-Platz 1, 37077 G\"ottingen, Germany}

\author{Kenji Watanabe} %
\address{Research Center for Electronic and Optical Materials, National Institute for Materials Science, 1-1 Namiki, Tsukuba 305-0044, Japan}

\author{Takashi Taniguchi} %
\address{Research Center for Materials Nanoarchitectonics, National Institute for Materials Science,  1-1 Namiki, Tsukuba 305-0044, Japan}

\author{Sabine Steil} 
\address{I. Physikalisches Institut, Georg-August-Universit\"at G\"ottingen, Friedrich-Hund-Platz 1, 37077 G\"ottingen, Germany}

\author{Daniel Steil} %
\address{I. Physikalisches Institut, Georg-August-Universit\"at G\"ottingen, Friedrich-Hund-Platz 1, 37077 G\"ottingen, Germany}

\author{R. Thomas Weitz} %
\address{I. Physikalisches Institut, Georg-August-Universit\"at G\"ottingen, Friedrich-Hund-Platz 1, 37077 G\"ottingen, Germany}
\address{International Center for Advanced Studies of Energy Conversion (ICASEC), University of Göttingen, Göttingen, Germany}

\author{Stephan Hofmann} 
\address{Department of Engineering, University of Cambridge, Cambridge CB3 0FA, U.K.}

\author{G.~S.~Matthijs~Jansen} %
\address{I. Physikalisches Institut, Georg-August-Universit\"at G\"ottingen, Friedrich-Hund-Platz 1, 37077 G\"ottingen, Germany}

\author{Samuel Brem} 
\address{Fachbereich Physik, Philipps-Universit{\"a}t, 35032 Marburg, Germany}

\author{Ermin Malic} 
\address{Fachbereich Physik, Philipps-Universit{\"a}t, 35032 Marburg, Germany}

\author{Marcel Reutzel} \email{marcel.reutzel@phys.uni-goettingen.de}%
\address{I. Physikalisches Institut, Georg-August-Universit\"at G\"ottingen, Friedrich-Hund-Platz 1, 37077 G\"ottingen, Germany}

\author{Stefan Mathias} \email{smathias@uni-goettingen.de}%
\address{I. Physikalisches Institut, Georg-August-Universit\"at G\"ottingen, Friedrich-Hund-Platz 1, 37077 G\"ottingen, Germany}
\address{International Center for Advanced Studies of Energy Conversion (ICASEC), University of Göttingen, Göttingen, Germany}

\begin{abstract}

The energy landscape of optical excitations in mono- and few-layer transition metal dichalcogenides (TMDs) is dominated by optically bright and dark excitons. These excitons can be fully localized within a single TMD layer, or the electron- and the hole-component of the exciton can be charge-separated over multiple TMD layers. Such intra- or interlayer excitons have been characterized in detail using all-optical spectroscopies, and, more recently, photoemission spectroscopy. In addition, there are so-called hybrid excitons whose electron- and/or hole-component are delocalized over two or more TMD layers, and therefore provide a promising pathway to mediate charge-transfer processes across the TMD interface. Hence, an in-situ characterization of their energy landscape and dynamics is of vital interest.  In this work, using femtosecond momentum microscopy combined with many-particle modeling, we quantitatively compare the dynamics of momentum-indirect intralayer excitons in monolayer WSe$_2$ with the dynamics of momentum-indirect hybrid excitons in heterobilayer WSe$_2$/MoS$_2$, and draw three key conclusions: First, we find that the energy of hybrid excitons is reduced when compared to excitons with pure intralayer character. Second, we show that the momentum-indirect intralayer and hybrid excitons are formed via exciton-phonon scattering from optically excited bright excitons. And third, we demonstrate that the efficiency for phonon absorption and emission processes in the mono- and the heterobilayer is strongly dependent on the energy alignment of the intralayer and hybrid excitons with respect to the optically excited bright exciton. Overall, our work provides microscopic insights into exciton dynamics in TMD mono- and bilayers.


\end{abstract}

\maketitle



\section{Introduction}

Exfoliated and artificially stacked monolayers of transition metal dichalcogenides (TMDs) have been shown to be a highly tuneable material platform for exploring optical excitations and correlated interactions on the atomic scale~\cite{Jin18natnano,Rivera18natnano,Kenness21natphys,Wilson21nat,Regan22natrevmat,Perea22apl,Raja18nanolett}. After first experiments identified the transition from an indirect to a direct semiconductor when exfoliating bulk TMDs to the monolayer limit~\cite{Mak10prl,Splendiani10nanolett}, subsequent studies characterized the strong exciton response to an optical excitation~\cite{Wang18rmp,Chernikov14prl,He14prl,Ye14nat}. Importantly, these pioneering all-optical experiments are for the most part only sensitive to radiative recombination processes within the light-cone~\cite{Yu15prl,Brem20nanolettsideband}. However, in addition to these bright excitons, it is meanwhile well-known that also tightly bound dark excitons contribute to the energy landscape of excitons~\cite{Zhang15prl,Christiansen17prl,Selig16natcom,Malic18prm,Deilmann192dmat,Selig182Dmat}. For example, for momentum-indirect dark excitons in a monolayer TMD, the exciton's electron and hole component can reside in different valleys of the TMD Brillouin zone (figure~1(c, e)). In the case that TMD homo- and heterostructures are assembled by more than one layer, the situation can become even more complex: The twist-angle between two neighbouring TMD layers can be varied from 0$\degree$ to 60$\degree$, and, in this way, the high-symmetry points of the Brillouin zones are offset in momentum space. Hence, so-called interlayer excitons for which the electron- and the hole-component are charge-separated across the interface are typically momentum-indirect and thus optically dark (figure~1(d, f))~\cite{Nayak17acsnano}. 

\begin{figure}[bt!]
    \centering
    \includegraphics{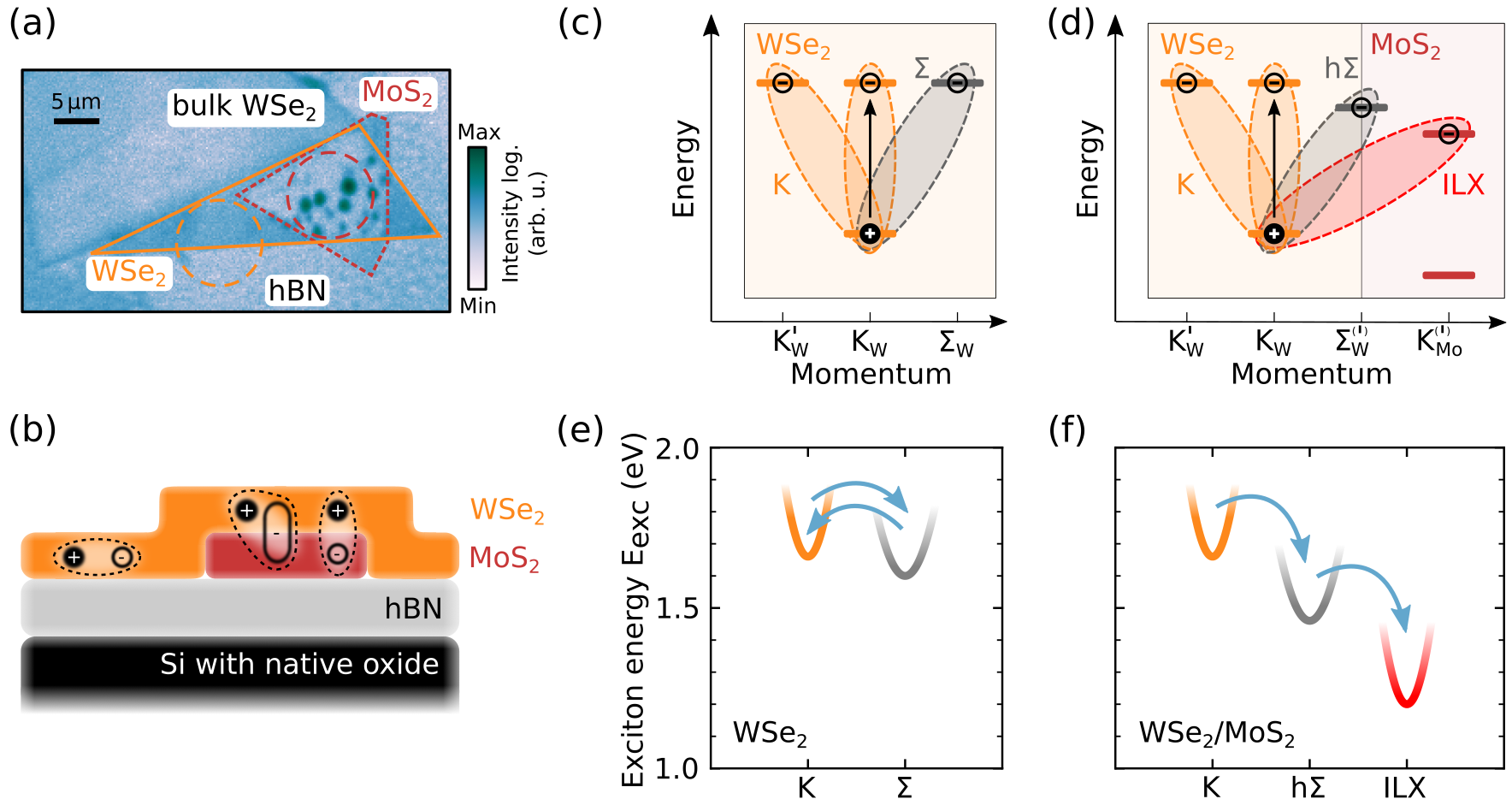}
    \caption{Real-space sample structure and energy landscape of monolayer WSe$_2$ and heterobilayer WSe$_2$/MoS$_2$.
    (a) Real-space resolved photoemission image of the monolayers WSe$_2$ (orange polygon) and MoS$_2$ (dark red dashed polygon). Femtosecond momentum microscopy experiments of the monolayer WSe$_2$ and the heterobilayer WSe$_2$/MoS$_2$ are performed in the regions-of-interest indicated by the orange and dark red circles, respectively.
    (b) Schematic sketch of the TMD heterostructure after excitation with a 1.7~eV light pulse. Intralayer excitons are formed in the WSe$_2$ monolayer and the WSe$_2$/MoS$_2$ heterobilayer. In the WSe$_2$/MoS$_2$ heterostructure, the energetically most stable state is the interlayer exciton where the electron and the hole component are localized in the WSe$_2$ and the MoS$_2$ layer, respectively. In addition, hybrid excitons are formed for which the exciton's hole is fully localized in the WSe$_2$ layer and the exciton's electron has probability density in the WSe$_2$ and the MoS$_2$ layer.
    (c,d) Single-particle energy landscape of monolayer WSe$_2$ and heterobilayer WSe$_2$/MoS$_2$. The filled bars indicate the single-particle valence and conduction band extrema at the Brillouin zone's high-symmetry points. The ovals indicate the Coulomb correlation between the single-particle electrons and holes, and the letters label the electron-hole pairs, as introduced in the main text.
    (e,f) Energy landscape of excitons in the WSe$_2$ monolayer  and the WSe$_2$/MoS$_2$ heterobilayer. The energies E$_{\rm exc}$ are obtained in experiment, and the blue arrows indicate the dominant scattering pathways.
    }
\end{figure}

First direct access to these dark excitons has been provided in optical-pump-midinfrared-probe experiments~\cite{Poellmann15natmat,Merkl19natmat}. With the development of high-repetition rate extreme ultraviolet light sources~\cite{Keunecke20timeresolved,puppin_time-_2019,Heyl12jpb,Li16rsi}, time- and angle-resolved photoelectron spectroscopy (trARPES)~\cite{Sobota21rmp,Bovensiepen12book} has become applicable to probe the energetics and dynamics of bright and dark excitons. First experiments focused onto the ultrafast exciton and charge carrier dynamics in semiconducting bulk~\cite{Bertoni16prl,Wallauer16apa,Hein16prb,Wallauer20prb,Dong20naturalsciences} and wafer-scale~\cite{Grubisic15nanolett,Liu19prl,Lin22prb,Lee21nanolett,Aeschlimann20sciadv} TMDs. More recently, empowered by the development of time-of-flight momentum microscopes~\cite{kromker_development_2008,medjanik_direct_2017}, trARPES has been successfully applied to probe the formation dynamics of momentum-indirect intralayer excitons~\cite{Madeo20sci,Wallauer21nanolett} and the valley depolarization dynamics~\cite{kunin23prl} in exfoliated monolayer TMDs. In addition, the cascaded exciton transition from bright intralayer to dark interlayer excitons has been quantified in an artificially stacked TMD heterobilayer~\cite{Schmitt22nat,Bange23arxiv}. 

These seminal trARPES experiments of exfoliated TMDs all focus on excitons that are either of full intra- or interlayer character: The exciton's electron and hole either reside both in a single TMD layer, or are charge-separated between both layers. Until today, no trARPES experiment has focused on the case where hybridization between the neighbouring TMD layers leads to the formation of a new type of exciton where either the electron-, the hole-, or the electron- and the hole-component are delocalized between both layers (figures~1(b)). Importantly, these so-called hybrid excitons are of great interest due to their potential to mediate charge transfer across a TMD interface~\cite{Wang17prb,Meneghini22naturalsciences,Schmitt22nat,Zimmermann21acsnano,Kunstmann18natphys,Policht23arxiv,Wallauer20prb}. Moreover, they have a high oscillator strength and a sensitivity to external electrical fields due to their partial intra- and interlayer character~\cite{Alexeev19nat,Tran19nat,Seyler19nat,Jin19nat,Lindlau18natcom,Deilmann192dmat,Berghauser18prb,Tagarelli23natpho}.

In this article, we study the impact of interlayer hybridization on the energy landscape of excitons and the resulting femto- to picosecond dynamics. In a joint theory-experiment effort, we directly compare the layer-localized $\Sigma$ exciton in monolayer WSe$_2$ with the hybrid h$\Sigma$ exciton in heterobilayer WSe$_2$/MoS$_2$ (figure~1(c-f)). For both excitons, the hole-component to the exciton is fully layer-localized and found in the K$_{\rm W}$ valley valence band maximum (VBM) of WSe$_2$. In contrast, the exciton's electron, which resides in the $\Sigma$ valley of the conduction band, is either fully layer localized (WSe$_2$) or has a significant degree of interlayer hybridization (WSe$_2$/MoS$_2$). First, we experimentally quantify the bright and dark exciton energies E$_{\rm exc}^i$ of the correlated two-particle energy landscape. Second, we identify exciton-phonon scattering as the dominating mechanism for the formation of intralayer $\Sigma$ and hybrid h$\Sigma$ excitons. And third, we show that the sub-ps exciton thermalization dynamics are significantly affected by the energy alignment of the $\Sigma$ and h$\Sigma$ exciton with respect to the optically excited exciton.


\section{Bright and dark excitons in WSe$_2$ and WSe$_2$/MoS$_2$}
\label{sec:theory}

The major goal of this manuscript is the identification of hybrid excitons in the trARPES experiment and the evaluation of their impact on the exciton relaxation dynamics. Therefore, we describe the low energy landscape of excitons as calculated by solving the Wannier equation~\cite{Meneghini22naturalsciences,brem2020hybridized,Perea22apl,Brem18scirep,Selig182Dmat,Deilmann192dmat}. All relevant excitations are summarized in the energy diagrams in figure~1(c-f).

If the monolayer WSe$_2$ is pumped with 1.7~eV photons, bright A1s excitons of WSe$_2$ are resonantly excited (black arrow in figure~1(c)). We label this exciton as K as its electron and hole component reside in the K$_{\rm W}$ valley conduction band minimum (CBM) and VBM, respectively. Note that throughout the manuscript, we do not differentiate between K excitons that are bound at K$_{\rm W}$ or K$^\prime_{\rm W}$ valleys, because we cannot distinguish them in experiment. In addition to the bright A1s exciton, two distinct momentum-indirect dark excitons can be formed in a subsequent scattering process: On the one hand, there are momentum-indirect dark excitons for which the electron- and the hole-component are found in the $\Sigma_{\rm W}$ and the K$_{\rm W}$ valleys, respectively; these excitons are labelled as $\Sigma$ throughout our work (figure~1(c, e)). On the other hand, momentum-indirect excitons for which the electron- and the hole-component are momentum-offset and found in the K$_{\rm W}$ or K$^\prime_{\rm W}$ valleys can be formed. Importantly, while it is expected that these momentum-indirect excitons are energetically favorable over the optically excited bright excitons~\cite{Berghauser18prb,Lindlau18natcom,Selig182Dmat}, where the electron- and the hole-component are bound in the same K$_{\rm W}$ or K$^{\prime}_{\rm W}$ valley, within the energy resolution of the photoemission experiment, it is not possible to separate these excitons. Hence, we label both of them with K throughout the manuscript.

In the WSe$_2$/MoS$_2$ heterostructure, 1.7~eV photons can be used to resonantly excite the A1s exciton in WSe$_2$ (black arrow in figure~1(d)). A similar resonance frequency for the optical excitation of WSe$_2$ A1s excitons in the mono- and the heterobilayer region can be rationalized based on the fact that wavefunction hybridization at the K$_{\rm W}$ valleys is negligible~\cite{Wang17prb,Kunstmann18natphys,brem2020hybridized,Karni19prl,Zhu17nanolett}. In addition, because the optical excitation of A1s excitons of monolayer MoS$_2$ would require at least 1.9~eV light-pulses~\cite{Li14prb,Bange23arxiv,Zimmermann21acsnano,Zhu17nanolett}, the MoS$_2$ monolayer is not directly excited. The optically excited WSe$_2$ A1s excitons can decay to form momentum-indirect excitons, where the electron- and the hole-component are momentum-offset between the K$_{\rm W}$ and K$^\prime_{\rm W}$ valleys. Because we cannot differentiate between these momentum-indirect excitons and the bright A1s excitons in the photoemission experiment, as in the case of the WSe$_2$ monolayer, we label both these bright and momentum-indirect excitons with K. More interestingly for our study, momentum-indirect hybrid h$\Sigma$ excitons are formed for which the exciton's hole and electron can be found in the K$_{\rm W}$ and the $\Sigma^{\left(\prime\right)}$ valley, respectively: Here, the $\Sigma^{\left(\prime\right)}$ valley conduction bands of the WSe$_2$ and MoS$_2$ are hybridized, and, in consequence, the h$\Sigma$ exciton can be described in the hybrid exciton basis with approximately 30\% intra- and 70\% interlayer character~\cite{Schmitt22nat}. Note that the hybrid character of this exciton's electron component is indicated by the letter 'h' in the abbreviation. With this hybrid character of the h$\Sigma$ excitons, interlayer charge transfer is strongly favored and interlayer excitons (ILX) can be formed~\cite{Schmitt22nat,Meneghini22naturalsciences}. For ILX, because of the negligible hybridization of the wavefunctions at the WSe$_2$ K$^{\left(\prime\right)}_{\rm W}$ and MoS$_2$ K$^{\left(\prime\right)}_{\rm Mo}$ valleys, the exciton's electron and hole component are again fully localized in the monolayers (figure~1(d)).


\section{Momentum microscopy: Energy landscape of excitons}

In this section, we experimentally quantify the exciton energies of the bright and dark excitons in the WSe$_2$ monolayer and the WSe$_2$/MoS$_2$ heterolayer. All energies are summarized in table~\ref{table:table}.

\subsection{Femtosecond momentum microscopy of exfoliated TMDs}

In order to establish the most direct experimental comparison of the exciton energy landscape of the WSe$_2$ monolayer and the twisted WSe$_2$/MoS$_2$ heterobilayer, we fabricate a single sample that contains distinct areas in which the monolayer WSe$_2$ flake and the WSe$_2$/MoS$_2$ heterostructure are found (figure~1(a)). We have chosen a doped silicon wafer with a native oxide layer as the substrate for the heterostructure, because it ensures high quality trARPES data that is not affected by sample charging (figure~1(b)). Moreover, the TMD flakes are stacked on 20-30~nm hexagonal boron nitride~\cite{Taniguchi07jcg} for best interface quality~\cite{Ulstrup19apl}. Before the trARPES experiments, the sample is annealed for 1~h to 670~K. In the real-space-resolved photoemission electron microscopy image in figure~1(a), the boundaries of the WSe$_2$ and MoS$_2$ monolayers are traced by orange and dark red (dashed) polygons, respectively. The twist-angle of the heterostructure is quantified to 9.8$\pm$0.8$^\circ$~\cite{Schmitt22nat}.

Time-resolved photoemission spectroscopy of exfoliated mono- and heterobilayers becomes possible by using our setup for femtosecond momentum microscopy (cf. refs.~\cite{Keunecke20timeresolved,Keunecke20prb,Li22prb}). We resonantly excite the bright A1s exciton of WSe$_2$ with 1.7~eV 40~fs light pulses ($s$-polarized). Photoemission from the occupied band structure and the excitons is induced with  time-delayed 26.5~eV 20~fs light pulses ($p$-polarized) that are created in a table-top 500~kHz repetition rate high-harmonic generation beamline~\cite{Keunecke20timeresolved,Duvel22nanolett,jansen_efficient_2020}. For each delay between the pump and the probe laser pulse, the time-of-flight momentum microscope (ToF-MM, Surface Concept GmbH) collects three-dimensional data cubes that contain information on the two in-plane momenta $k_x$ and $k_y$ and the energy $E$ of the detected photoelectrons~\cite{medjanik_direct_2017,Keunecke20timeresolved}. Notably, by inserting an aperture into the real-space plane of the microscope, the experiment is sensitive to the exciton dynamics in a sample region with a diameter of 10~$\mu$m (orange and red circles in figure~1(a))~\cite{Schmitt22nat,Madeo20sci,Wallauer21nanolett,kunin23prl}. 


The momentum microscopy experiment can be used to directly characterize and compare the occupied band structure of monolayer WSe$_2$ and heterobilayer WSe$_2$/MoS$_2$ (figure~2). In both sample areas, the spin-split valence bands at the WSe$_2$ K$_{\rm W}$ valley can be identified (labelled (1) and (2) in the magenta energy-distribution-curves (EDCs)). In the heterobilayer, in addition, the MoS$_2$ valence band is observed at E-E$_{\rm VBM}=0.94\pm0.05$~eV (labelled (3) in figure~2(b)). While the dispersion of the WSe$_2$ valence bands are similar at the K$_{\rm W}$ valleys of the mono- and the heterobilayer, it is strikingly different at the $\Gamma$ valley: In the monolayer, only a single band at the $\Gamma$ valley is observed (horizontal arrow, figure~2(a)). In contrast, in the heterobilayer, we find two energetically separated bands that are a clear indication for interlayer hybridization (two horizontal arrows in figure~2(b))~\cite{Wilson17sciadv,Stansbury21sciadv,Jones21twodmat}.


\begin{figure}[tb]
    \centering
    \includegraphics{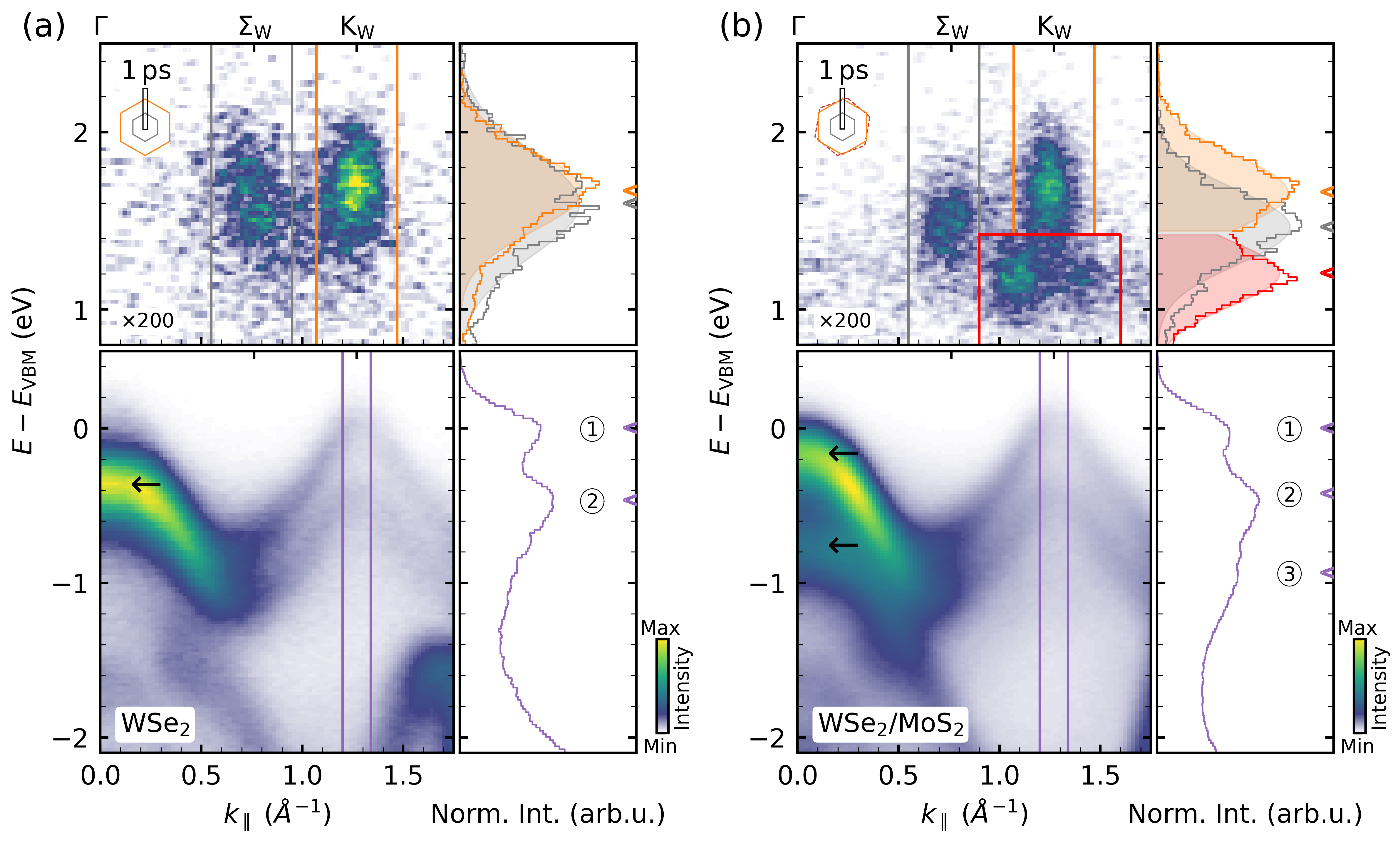}
    \caption{Energy- and momentum-resolved photoemission spectra of monolayer WSe$_2$ and heterobilayer WSe$_2$/MoS$_2$ along the $\Gamma$-K$_{\rm W}$ direction (inset).
    (a) The occupied valence band structure of monolayer WSe$_2$ is dominated by a single-band at the $\Gamma$ valley (black horizontal arrow) and the spin-split valence bands at the K$_{\rm W}$ valley (labelled (1) and (2) in the EDC). At a pump-probe delay of 1~ps and a photoemission energy of $\approx$1.7~eV, exciton photoemission yield from intralayer K (orange) and $\Sigma$ (grey) excitons is detected. 
    (b) The occupied valence band structure of heterobilayer WSe$_2$/MoS$_2$ is characterized by the WSe$_2$ spin-split valence bands at the K$_{ \rm W}$ valley (labelled (1) and (2) in the EDCs) and the valence band of MoS$_2$ (3), for which the spin-splitting is not resolved. At the $\Gamma$ valley, interlayer hybridization leads to the formation of two bands (horizontal black arrows). At a pump-probe delay of 1~ps, exciton photoemission yield is detected from the intralayer K excitons (orange), hybrid h$\Sigma$ excitons (grey) and ILX (red). 
    The photoemission energy from hybrid h$\Sigma$ exciton's is significantly reduced when compared to the intralayer K (monolayer and heterobilayer) and $\Sigma$ (monolayer) excitons. The energy-momentum cuts are obtained by summation of all six measured $\Gamma$-K$_{\rm W}$ ($\Gamma$-K$^{\prime}_{\rm W}$) directions.
    }
\end{figure}

\begin{figure}[tb]
    \centering
    \includegraphics{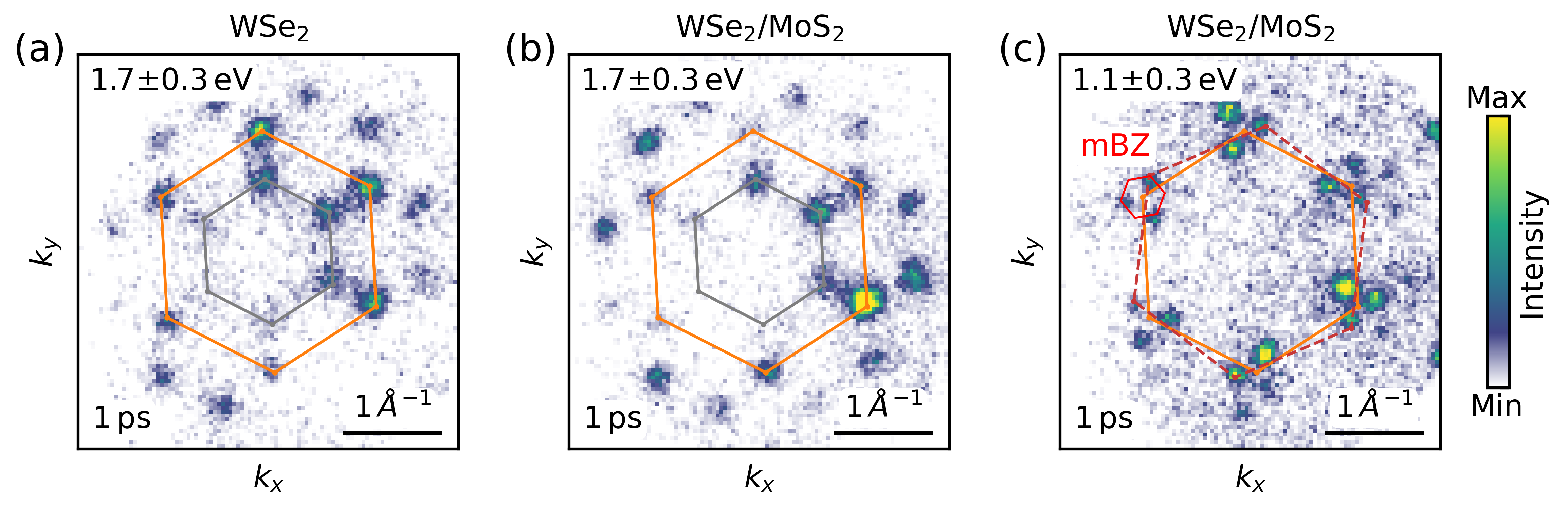}
    \caption{Momentum-resolved photoemission maps at the energies of the intralayer, hybrid and interlayer excitons in monolayer WSe$_2$ and heterobilayer WSe$_2$/MoS$_2$. The pump-probe delay and the photoemission energy with respect to the WSe$_2$ VBM are given in the figure. All momentum maps are integrated in an energy window $\pm$0.3~eV around the center energy.
    (a) In the WSe$_2$ monolayer, the exciton photoemission momentum fingerprints of intralayer K (orange) and $\Sigma$ (grey) show spectral weight at the K$^{\left(\prime\right)}_{\rm W}$ and $\Sigma^{\left(\prime\right)}_{\rm W}$ valleys of the WSe$_2$ hexagonal Brillouin zone.
    (b) In the WSe$_2$/MoS$_2$ heterobilayer, exciton photoemission from the intralayer K (orange) and the hybrid h$\Sigma$ (grey) excitons is found at the K$^{\left(\prime\right)}_{\rm W}$ and $\Sigma^{\left(\prime\right)}_{\rm W}$ valleys of the WSe$_2$ hexagonal Brillouin zones. 
    (c) The momentum-misalignment of the Brillouin zones of WSe$_2$ (orange hexagon) and MoS$_2$ (dark red dashed hexagon) leads to a complex photoemission signature from ILX that can be described within the moiré mini Brillioun zone (mBZ, red hexagon).
    }
\end{figure}

\subsection{Spectroscopy of excitons in monolayer WSe$_2$ and heterobilayer WSe$_2$/MoS$_2$}

Having identified these distinct spectroscopic signatures of the electronic band structure that allow the direct discrimination of the mono- and the heterobilayer region in the trARPES experiment, in the next step, we probe the energetics of the optical excitations of the WSe$_2$ monolayer. Therefore, optical pump pulses with a center energy of 1.7~eV are used in order to be resonant to the WSe$_2$ A1s exciton (fluence:  280~$\mu$J/cm$^{-2}$, exciton density: 5.4$\times 10^{12}$~excitons/cm$^{-2}$). We find spectral weight above the valence bands that we attribute to photoelectrons being emitted from excitons (figure~2(a)). Specifically, at a pump-probe delay of 1~ps, photoemission yield is detected at the K$_{\rm W}$ and the $\Sigma_{\rm W}$ valleys of the WSe$_2$ Brillouin zone (orange and grey hexagon in figure~(3a), respectively). These spectral signatures can be attributed to photoemission from intralayer K and $\Sigma$ excitons, consistent with earlier reports~\cite{Madeo20sci,Wallauer21nanolett}. 


When applying similar excitation conditions to the WSe$_2$/MoS$_2$ heterobilayer region, much richer spectral signatures are detected in the momentum microscopy experiment: A complex distribution of photoemission intensity is detected above the WSe$_2$ valence bands in an energy window ranging from 0.9~eV to 1.9~eV (figure~2(b), 1~ps pump-probe delay). The momentum-map centered at an energy of 1.7~eV above the WSe$_2$ VBM shows spectral weight at the momenta of the K$_{ \rm W}$ and the $\Sigma_{ \rm W}$ valleys that we attribute to photoemission signal from intralayer K and hybrid h$\Sigma$ excitons (figure~3(b), orange and grey hexagon, respectively). At a lower energy (figure~3(c), E-E$_{\rm VBM}\approx1.1$~eV), we find a complex momentum structure of the photoemission intensity that can be attributed to ILX: Spectral weight is detected at the K$_{\rm Mo}$ valley and the additional $\kappa$ valleys that are described within the moiré mini Brillouin zone (mBz, red hexagon). This hallmark of the moiré superlattice imprinted onto the exciton photoemission signature from the interlayer ILX excitons is discussed in detail in ref.~\cite{Schmitt22nat}.


\subsection{Energy landscape of excitons in WSe$_2$ and WSe$_2$/MoS$_2$}

Having identified the major photoemission spectral signatures of excitons, we aim to experimentally quantify the energy landscape of bright and dark excitons in the mono- and the heterobilayer sample. For this, first, we have to discuss at which energy the photoemission experiment detects single-particle photoelectrons that have initially been bound in the correlated exciton state. In the process of photoemission from an exciton, the Coulomb correlation between the exciton's electron and hole is broken at the cost of the exciton binding energy. As described by Weinelt \textit{et al.}~\cite{Weinelt04prl} and others~\cite{Zhu14jpcl,Perfetto16prb,Rustagi18prb,Christiansen19prb,Tanimura19prb,Bange23arxiv,Man21sciadv,Bennecke23arxiv,Dong20naturalsciences}, the single-particle photoelectron from an exciton will be detected one exciton energy E$_{\rm exc}^i$ above the energy of the single-particle band where the former hole contribution to the exciton remains in the sample, i.e., at 
\begin{equation}
    E_{\rm elec}=E_{\rm hole}+E^i_{\rm exc}+\hbar\omega
    \label{eq:energyconser}
\end{equation}
with $E_{\rm hole}$ and $E_{\rm elec}$ as the  single-particle energy of hole and the electron-state, respectively; furthermore with $E^i_{\rm exc}$ as the exciton energy and $\hbar\omega$ as the probe photon energy. Importantly, Eq.~\ref{eq:energyconser} sets the energy of the WSe$_2$ VBM at the K$_{\rm W}$ valley as the natural reference point of the present experiment, because the single-particle hole of all probed excitons remains in this valley once photoemission has occurred (cf. figure~1(c,d)). 

Hence, we can quantify the exciton energy $E_{\rm exc}^i$ of all contributing intralayer, hybrid and interlayer excitons by analysing the EDCs taken at the K$_{\rm W}$, the $\Sigma_{\rm W}$ and the K$_{\rm Mo}$ ($\kappa$) valleys (figure~2). Following Eq.~\ref{eq:energyconser}, we then calculate the energy difference between each exciton photoemission signal and the energy of the K$_{\rm W}$ valley VBM. All exciton energies $E_{\rm exc}^i$ quantified for the WSe$_2$ mono- and the WSe$_2$/MoS$_2$ heterobilayer are summarized in table~\ref{table:table}. 

Having experimentally measured the exciton energies $E_{\rm exc}^i$, we are now in the position to systematically compare those values. First, we find that, within the experimental error, the intralayer K exciton energy is comparable in the mono- and the heterobilayer region (1.67$\pm$0.05~eV vs. 1.66$\pm$0.05~eV). For WSe$_2$/MoS$_2$, this result strongly supports the expectation that excitons, where the electron- and hole-component are localized in the K$_{\rm W}$ valleys, are not affected by interlayer hybridization~\cite{brem2020hybridized,Wang17prb,Kunstmann18natphys}. In contrast, we find that the hybrid character of the h$\Sigma$ excitons indeed leads to a renormalization of the exciton energy: The interlayer hybridization leads to a reduction of the exciton energy from 1.60$\pm$0.05~eV ($\Sigma$ exciton, monolayer WSe$_2$) to 1.46$\pm$0.05~eV (h$\Sigma$ exciton, heterobilayer WSe$_2$/MoS$_2$). This observation is a direct experimental verification of seminal photoluminescence experiments that reported a changed emission energy for hybrid excitons, if the degree of hybridization is controlled, e.g., by the twist angle~\cite{Merkl20natcom,Kunstmann18natphys,Zande14nanolett,Liu14natcom}.

\begin{table}[tb!]
\caption{Summary of the experimentally quantified exciton energies E$_{\rm exc}^{\rm i}$ in monolayer WSe$_2$ and heterobilayer WSe$_2$/MoS$_2$.}
\begin{center}
\begin{tabular}{ |c||c|c|} 
 \hline
        & \,\, monolayer WSe$_2$ \,\, & \,\, heterobilayer WSe$_2$/MoS$_2$ \,\, \\
 \hline
 \hline
    \,\, $E_{\rm exc}^{\rm K}$ (eV) \,\, &  1.67$\pm$0.05 & 1.66$\pm$0.05\\ 
    $E_{\rm exc}^{\rm \Sigma}$ (eV) & 1.60$\pm$0.05 & -\\ 
    $E_{\rm exc}^{\rm h\Sigma}$ (eV) &  - & 1.46$\pm$0.05  \\ 
    $E_{\rm exc}^{\rm ILX}$ (eV) & - & 1.20$\pm$0.05\\ 
 \hline                 
\end{tabular}
\end{center}
\label{table:table}
\end{table}


\section{Ultrafast exciton dynamics in WSe$_2$ and {WSe$_2$/MoS$_2$}}

Having characterized the energy landscape of excitons, we now turn to the exciton scattering dynamics. Specifically, the goal of the analysis is to pinpoint the impact of the intralayer $\Sigma$ and hybrid h$\Sigma$ excitons on the relaxation dynamics. Therefore, first, we compare the experimentally measured exciton dynamics with microscopic many-particle calculations that allow us to identify exciton-phonon scattering as the dominant mechanism for intervalley thermalization and the formation of the intralayer $\Sigma$ and hybrid h$\Sigma$ excitons. Second, we focus on the sub-ps dynamics in order to elucidate the impact of the different exciton energies of the intralayer $\Sigma$ and the hybrid h$\Sigma$ exciton on the exciton relaxation cascade.

\begin{figure}[bt!]
    \centering
    \includegraphics{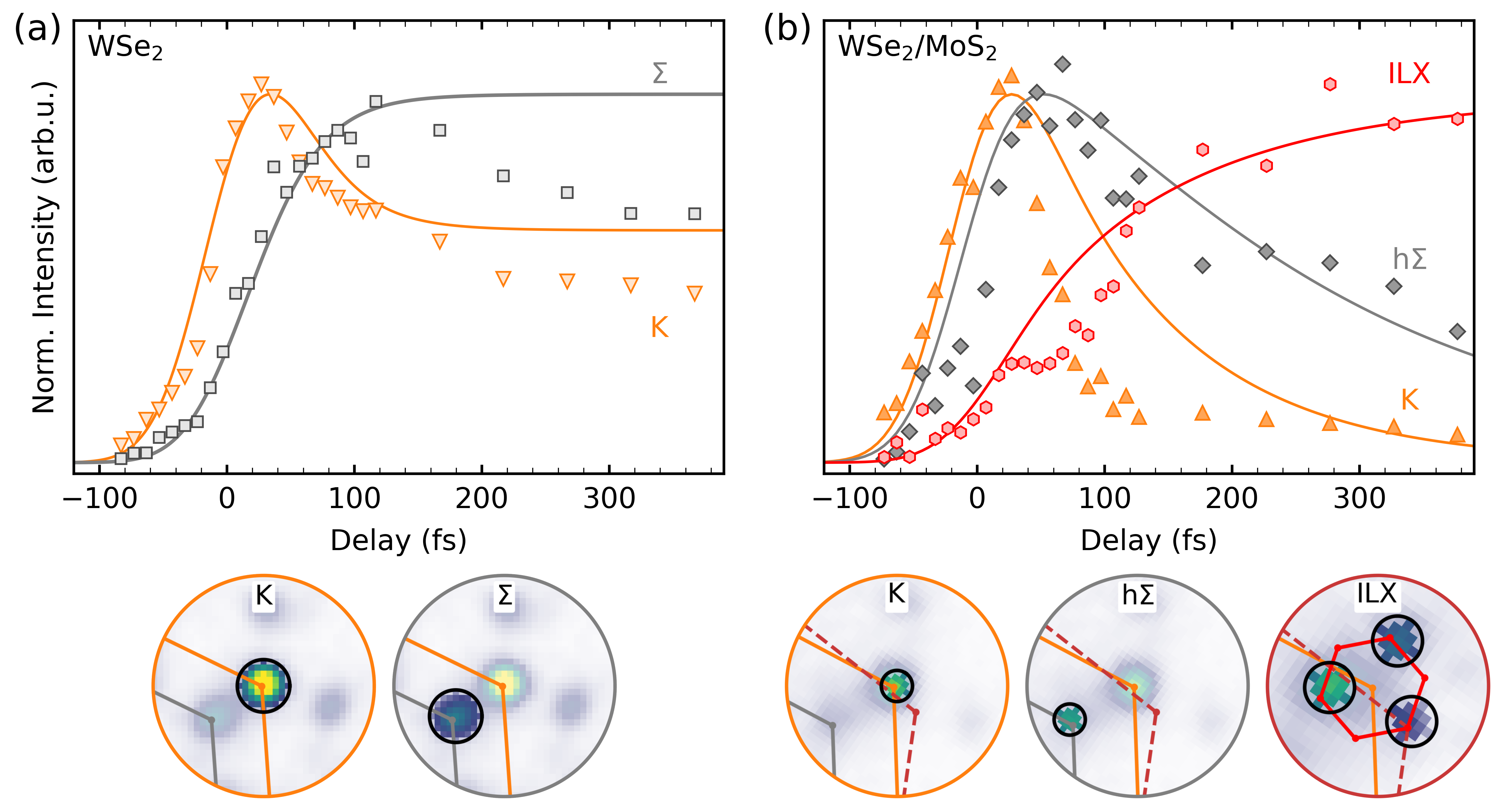}
    \caption{Direct comparison of the ultrafast exciton dynamics in (a) monolayer WSe$_2$ and (b) heterobilayer WSe$_2$/MoS$_2$. The symbols show the pump-probe  delay-dependent photoemission yield from excitons, and the solid lines are based on microscopic calculations. The exciton photoemission yield is filtered in the black-circled regions-of-interest indicated in the momentum-maps.
    }
\end{figure}


\subsection{Exciton-phonon scattering: Formation of intralayer $\Sigma$ and hybrid h$\Sigma$ excitons}

The femtosecond momentum microscopy experiment provides direct access to the ultrafast dynamics of bright and dark excitons. The symbols in the main panels of figure~4 show the pump-probe delay-dependent photoemission yield from the intralayer K (orange), intralayer $\Sigma$ (grey), hybrid h$\Sigma$ (grey) and ILX (red) excitons. In addition, the femtosecond evolution of exciton occupation as calculated in our microscopic model is plotted as solid lines of the respective color.

In the WSe$_2$ monolayer (figure~4(a)), photoemission yield from the K exciton peaks at pump-probe delays close to 40~fs. Delayed to this process, we find that spectral weight from momentum-indirect $\Sigma$ excitons (grey) increases on a 100~fs timescale. This rise of the photoemission yield from the intralayer $\Sigma$ excitons is in agreement with momentum microscopy experiments by Mad\'{e}o \textit{et al.}~\cite{Madeo20sci} and Wallauer \textit{et al.}~\cite{Wallauer21nanolett}. Moreover, the rise time is also in agreement with our microscopic calculations that include exciton-light and exciton-phonon interaction. From the model, we therefore identify exciton-phonon scattering as the dominant mechanism for the formation of intralayer $\Sigma$ excitons. Note that for pump-probe delays >100~fs, the exciton occupation calculated in the model overestimates the experimentally measured photoemisison intensity from both excitons. The reason for this deviation is that the model does not include decay processes and thus overestimates the exciton occupation for longer pump-probe delays.

The pump-probe delay-dependent photoemission yield from excitons measured in the WSe$_2$/MoS$_2$ heterobilayer is shown in figure~4(b). We find a direct hierarchy in the onset of rising photoemission yield from the three types of excitons, i.e., the K excitons (orange), the hybrid h$\Sigma$ excitons (grey), and the ILX (red). Again, the dynamics is in excellent agreement with our microscopic calculations, such that we can identify exciton-phonon scattering as the dominant mechanism for the formation of hybrid h$\Sigma$ excitons.


\begin{table}[tb!]
\caption{Summary of the experimentally quantified rise and decay times of photoemission yield from excitons.}
\begin{center}
\begin{tabular}{ |c||c|c|} 
 \hline
                                                        & \,\, monolayer WSe$_2$ \,\,  & \,\, heterobilayer WSe$_2$/MoS$_2$ \,\, \\
 \hline
 \hline
    \,\, $\tau_{\rm fast}^{\rm K}$ (fs) \,\,      & 70$\pm$10              & 28$\pm$2 \\ 
    $\tau_{\rm slow}^{\rm K}$ (fs)                & 1500$\pm$200           & 1900$\pm$800 \\
\hline
    $\tau_{\rm WSe_2}^{\rm\Sigma}$ (fs)          & 1050$\pm$60            & - \\ 
    \,\, $\tau_{\rm WSe_2/MoS_2, fast}^{\rm h\Sigma}$ (fs) \,\,   & -             & 110$\pm$40 \\ 
    $\tau_{\rm WSe_2/MoS_2, slow}^{\rm h\Sigma}$ (fs)             & -             & 1600$\pm$800 \\ 
\hline
    \,\, $\tau_{\rm rise}^{\rm \Sigma}$ (fs) \,\,   & 36$\pm$3                   & - \\ 
    \,\, $\tau_{\rm rise}^{\rm h\Sigma}$ (fs) \,\,   & -                  & 34$\pm$3 \\ 
\hline
    \,\, $\tau_{\rm rise}^{\rm ILX}$ (fs) \,\,   & -                   & 120$\pm$20 \\ 
 \hline                 
\end{tabular}
\end{center}
\label{table:tabletimes}
\end{table}

\subsection{Monolayer WSe$_2$: Steady state of phonon emission and absorption processes}

\begin{figure}[bt]
    \centering
    \includegraphics{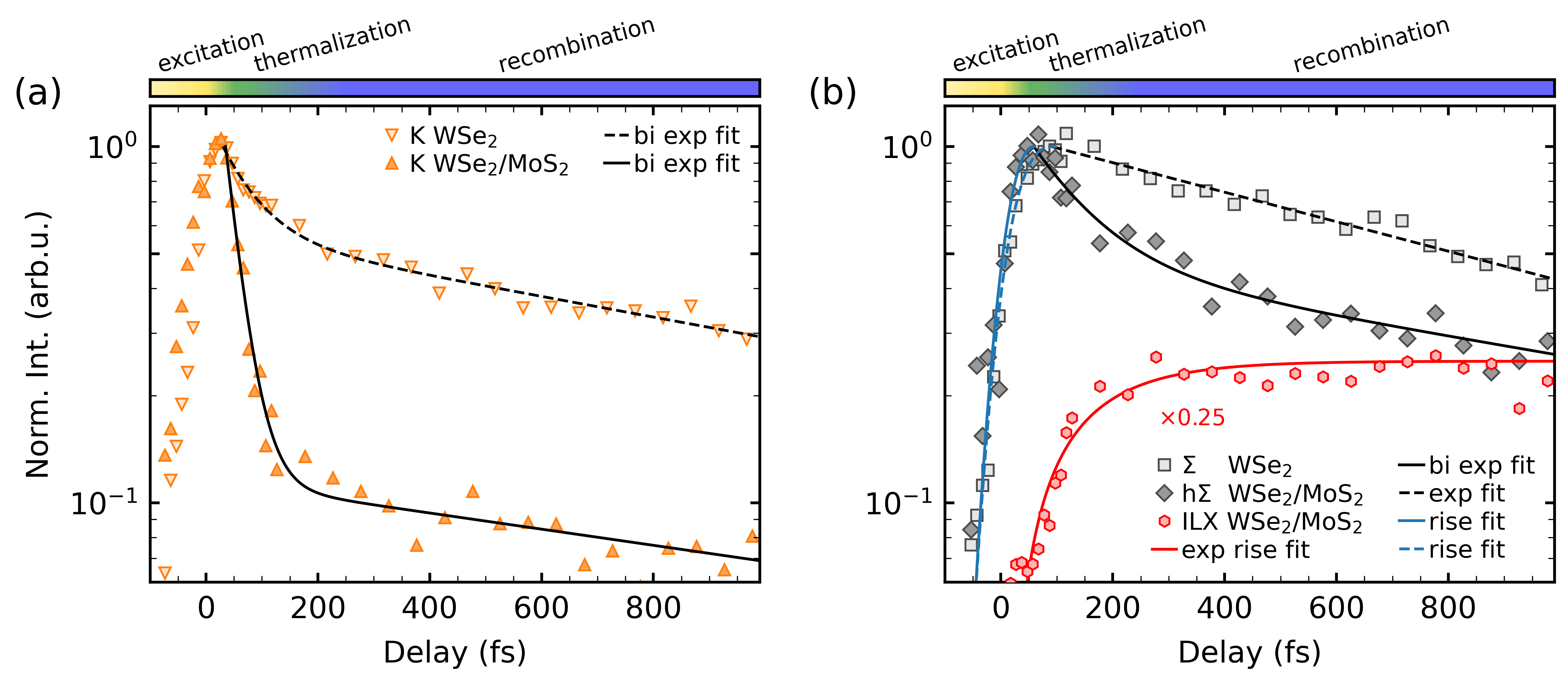}
    \caption{Quantitative analysis of the intralayer and hybrid exciton dynamics. In the yellow, green and blue shaded delay regions, excitons are optically excited, thermalize and radiatively recombine, respectively.
    (a) Pump-probe-delay evolution of photoemission signal from K excitons in monolayer WSe$_2$ (empty triangles) and heterobilayer WSe$_2$/MoS$_2$ (filled triangles); the dashed and solid lines are bi-exponential decay fits to the data. 
    (b) Quantitative evaluation of the formation and relaxation dynamics of intralayer $\Sigma$ excitons (empty squares, grey) and hybrid h$\Sigma$ excitons (filled squares, grey). The rise time of the excitons is evaluated with a one-level rate equation model (blue dashed and solid line, respectively). The decay dynamics is approximated by bi- and single-exponential fits, respectively. The rise time of the ILX (red circles) is fitted with a single exponential function (red line). All rise and decay constants are summarized in table~\ref{table:tabletimes}.
    }
\end{figure}

So far, we have found that the intervalley relaxation processes in monolayer WSe$_2$ and heterobilayer WSe$_2$/MoS$_2$ are phonon-mediated and lead to the formation of intralayer $\Sigma$ and hybrid h$\Sigma$ excitons, respectively. However, even though the intervalley relaxation proceeds via the same mechanism, we find distinctly different decay dynamics of the contributing intralayer and hybrid excitons on the sub-ps timescale (figure~5 and table~\ref{table:tabletimes}). In the following, we pinpoint the origin of the different dynamics to the relative energy alignment of the intralayer $\Sigma$ and hybrid h$\Sigma$ excitons with respect to the intralayer K excitons, which has direct impact on the efficiency of exciton-phonon scattering processes.

In monolayer WSe$_2$, the energy difference between the optically excited K and the intralayer $\Sigma$ exciton is $\Delta E_{\rm exc}^{\rm WSe_2} = E_{\rm exc}^{\rm K}-E_{\rm exc}^{\Sigma}=0.07\pm0.10$~eV. Hence, phonon emission and absorption processes with typical energies of 0.03~eV~\cite{Jin14prb} can efficiently create a steady state between both excitonic species, i.e., K$\rightleftharpoons\Sigma$, where the relative exciton occupation is given by degenerate Bose-Einstein distributions~\cite{Selig182Dmat}. In order to identify this steady state in the photoemission data, we plot the sub-ps pump-probe delay-dependent photoemission intensity from K and $\Sigma$ excitons on a logarithmic intensity axis in figure~5(a) and 5(b), respectively. As indicated by the yellow, green and blue marked areas, the sub-ps dynamics of K excitons can be divided into overall three characteristic timescales. On the sub-40-fs timescale (yellow, figure~5(a)), photoemission yield from K excitons increases due to the optical excitation. For increasing pump-probe delay, the decaying photoemission intensity can be approximated by a bi-exponential function. Here, the two decay times $\tau_{\rm WSe_2, fast}^{\rm K}=70\pm10$~fs and $\tau_{\rm WSe_2, slow}^{\rm K}=1.5\pm0.2$~ps describe the fast and the slow component of the dynamics that are dominant in the green and blue delay-regions, respectively. In direct comparison to time-resolved all-optical spectroscopies, the slow component $\tau_{\rm WSe_2, slow}^{\rm K}$  can be attributed to radiative recombination processes of the bright exctions~\cite{Zimmermann21acsnano,Zhu17nanolett,Cui14acsnano}. However, the interpretation of the fast time constant $\tau_{\rm WSe_2, fast}^{\rm K}$ is more complex (green): First, it contains information on the decay of the coherent exciton polarization towards the incoherent K exciton population~\cite{Wallauer21nanolett,Trovatello20natcom}, which is not the focus of our study. Second, and more importantly, $\tau_{\rm WSe_2, fast}^{\rm K}$ is dominant in the delay-window that is necessary to establish a steady state between phonon absorption and emission events transferring K into $\Sigma$ excitons and vice versa. In agreement with this assignment, we find that photoemission intensity from momentum-indirect $\Sigma$ excitons, which are formed due to the decay of K excitons, rises on the same timescale and peaks at 100~fs (yellow and green delay-region in figure~5(b)). For longer delays (blue delay-region), the $\Sigma$ exciton photoemission intensity decays and is described by a single-exponential fit with a decay time of $\tau_{\rm WSe_2}^{\rm \Sigma}=1.1\pm0.1$~ps. Notably, it has already been suggested that the decay of momentum-indirect $\Sigma$ excitons dominantly proceeds via phonon absorption processes that first transfer the $\Sigma$ excitons into bright K excitons that can, subsequently, decay in a radiative process~\cite{Yuan20natmat,Selig182Dmat,Brem20nanolettsideband}. Our analysis verifies this proposition, because the experimentally quantified decay time $\tau_{\rm WSe_2}^{\rm \Sigma}$ of dark $\Sigma$ excitons is in reasonable agreement with the radiative decay time $\tau_{\rm WSe_2, fast}^{\rm K}$ of K excitons (cf. table~\ref{table:tabletimes}).


\subsection{Heterobilayer WSe$_2$/MoS$_2$: Phonon emission processes and formation of interlayer excitons}

The situation is significantly different for the WSe$_2$/MoS$_2$ heterobilayer. The hybrid character of the h$\Sigma$ exciton leads to a reduction of its exciton energy and thus to an energy difference of $\Delta E_{\rm exc}^{\rm WSe_2/MoS_2}=E_{\rm exc}^{\rm K}-E_{\rm exc}^{\rm h\Sigma}=0.20\pm0.10$~eV with respect to the intralayer K exciton. Hence, accompanied by the emission of a phonon, optically excited K excitons can effectively be transferred to hybrid h$\Sigma$ excitons. However, the large energy difference $\Delta E_{\rm exc}^{\rm WSe_2/MoS_2}$ strongly suppresses backscattering events for which multiple phonon absorption processes would be required. The suppression of the backscattering channel in the heterobilayer in comparison to the monolayer is most evident by the much stronger reduction of photoemission intensity from K exctions to below 10\% after only 100~fs (green delay-region in figure~5(a)). Photoemission intensity from hybrid h$\Sigma$ excitons, on the other side, rises on the same timescale, which is in agreement with the interpretation of an initially fast transfer of excitons from K to h$\Sigma$ excitons (yellow and green delay-regions in figures~5(b)).

Once the hybrid h$\Sigma$ excitons are formed, they can relax to the ILX by subsequent exciton-phonon scattering processes within the MoS$_2$ layer. For the analysis of this process, we fit the picosecond evolution of photoemission intensity from hybrid h$\Sigma$ excitons with a biexponential function (figure~5(b)) and extract a fast h$\Sigma\rightarrow$ILX decay time of $\tau_{\rm WSe_2/MoS_2, fast}^{\rm h\Sigma}=110\pm40$~fs. Notably, this decay time is in excellent agreement with the rise time of photoemission intensity from ILX (figure~5(b), $\tau_{\rm WSe_2/MoS_2, rise}^{\rm ILX}=120\pm 20$~fs, red data points), being thus a strong indication that ILX excitons are directly formed from hybrid h$\Sigma$ excitons. Again, because of the large energy difference between the hybrid h$\Sigma$ exciton and the ILX, i.e., $\Delta E_{\rm exc}^{\rm ILX}=E_{\rm exc}^{\rm h\Sigma}-E_{\rm exc}^{\rm ILX}=0.26\pm0.10$~eV (cf. table~\ref{table:table}), phonon absorption processes are nearly fully suppressed and backscattering events from ILX to hybrid h$\Sigma$ excitons can also be excluded.

Finally, we note that our experimental observation on the efficiency of forward and backward scattering in the WSe$_2$ monolayer and the WSe$_2$/MoS$_2$ heterobilayer are fully consistent with our microscopic model. Specifically, for the WSe$_2$ monolayer, the calculations reproduce the establishment of a steady state K$\rightleftharpoons\Sigma$ (blue arrows in figure~1(e)). Notably, the model here finds that especially the momentum-indirect excitons, where the hole- and the electron-component are momentum-offset in the K$_{\rm W}$ and K$^\prime_{\rm W}$ valleys, contribute to the establishment of the steady state, as detailed, e.g., in refs.~\cite{Selig182Dmat,Brem20nanolettsideband}. In addition, for the WSe$_2$/MoS$_2$ heterobilayer, the microscopic model reproduces the negligible efficiency for backscattering as compared to the forward scattering leading to the formation of ILX (blue arrows in figure~1(f)).


\section{Conclusion}

In a joint experiment-theory effort, we have used femtosecond momentum microscopy and microscopic modelling to characterize the energy landscape of bright and dark excitons and the resulting scattering dynamics in monolayer WSe$_2$ and heterobilayer WSe$_2$/MoS$_2$. First, we find that interlayer hybridization in WSe$_2$/MoS$_2$ leads to a reduction of the energy of the optically dark hybrid exciton when compared to monolayer WSe$_2$, where the exciton is localized in the layer. Second, in direct comparison to microscopic modelling, we demonstrate that the hybrid h$\Sigma$ and the intralayer $\Sigma$ excitons are formed via exciton-phonon scattering. And third, we show that the relative efficiency of phonon absorption and emission processes are the dominant parameter for the different exciton dynamics in monolayer WSe$_2$ and heterobilayer WSe$_2$/MoS$_2$. Specifically, in monolayer WSe$_2$, we showed that the occupation of intralayer K and $\Sigma$ evolves into a steady state, i.e., K$\rightleftharpoons\Sigma$, and the overall exciton occupation decays via radiative processes of the bright K excitons. In contrast, in the WSe$_2$/MoS$_2$ heterobilayer, we found that the energy level alignment of the contributing excitons strongly suppresses phonon-absorption processes and leads to a true exciton cascade K$\rightarrow\Sigma\rightarrow$ILX with negligible contributions of backscattering. 



\section{ACKNOWLEDGEMENTS}

This work was funded by the Deutsche Forschungsgemeinschaft (DFG, German Research Foundation) - 432680300/SFB 1456, project B01, 217133147/SFB 1073, projects B07 and B10, and 223848855/SFB 1083, project B9. A.A. and S.H. acknowledge funding from EPSRC (EP/T001038/1, EP/P005152/1). A.A. acknowledges financial support by the Saudi Arabian Ministry of Higher Education. E. M. acknowledges support from the European Unions Horizon 2020 research and innovation programme under grant agreement no. 881603 (Graphene Flagship). K.W. and T.T. acknowledge support from the JSPS KAKENHI (Grant Numbers 20H00354, 21H05233 and 23H02052) and World Premier International Research Center Initiative (WPI), MEXT, Japan. 

\bibliography{bibtexfile}

\end{document}